# Fine structure in the α-decay of odd-even nuclei

K. P. Santhosh[a,*], Jayesh George Joseph and B. Priyanka

*School of Pure and Applied Physics, Kannur University, Payyanur Campus, Payyanur 670 327, India*

**Abstract**

Systematic study on α-decay fine structure is presented for the first time in the case of odd-even nuclei in the range $83 \leq Z \leq 101$. The model used for the study is the recently proposed Coulomb and proximity potential model for deformed nuclei (CPPMDN), which employs deformed Coulomb potential, deformed two term proximity potential and centrifugal potential. The computed partial half lives, total half lives and branching ratios are compared with experimental data and are in good agreement. The standard deviation of partial half-life is 1.08 and that for branching ratio is 1.21. Our formalism is also successful in predicting angular momentum hindered and structure hindered transitions. The present study reveals that CPPMDN is a unified theory which is successful in explaining alpha decay from ground and isomeric state; and alpha fine structure of even-even, even-odd and odd-even nuclei. Our study relights that the differences in the parent and daughter surfaces or the changes in the deformation parameters as well as the shell structure of the parent and daughter nuclei, influences the alpha decay probability.

[*]Tel: +919495409757; Fax: +914972806402

*Email address*: drkpsanthosh@gmail.com

PACS number(s): 23.60. +e; 24.10.−i; 27.80.+w; 27.90.+b

*Keywords*: Alpha decay, Fine Structure

## 1. Introduction

The ground state to ground state transitions in α-decay is only a channel among many possibilities of the decay. This inherently conceives an idea of decay to other states and possibly invites the term α-fine-structure. The α fine-structure was discovered by Rosenblum [1] since 1929 by measuring the range of emitted particle in air. It reveals the essential variation of particle emission probability for different states of daughter nucleus. The variation of intensity is likely due to structure incompatibilities between parent and daughter nucleus. There are two realizations, namely, the structure hindrance and the angular momentum hindrance; structure hindrance, means that states having dissimilar 'appearance' accompany decay with a hindrance. On the other hand, transitions may be impeded due to transferring of angular momentum, which makes it angular momentum hindered, called unfavored transitions. The interplay among these two effects leads to a spectrum in energy distribution of α-particles from a single element. Using this, the identification of states [2], persistence of shell closure [3], coexistence of nuclear shapes [4] and so on, can effectively be probed. Without still disclosing the intriguing nature, the α-decay, is now considered as an object which has wide applicability in nuclear structure study.

Although various theoretical attempts have been done on alpha fine structure, it is not yet qualitatively explained. Especially during the last few decades various existing theoretical models such as shell model, fission like model and cluster model have been used and modified so as to explain the fine structure of α-decay. In 1956, J. O. Rasmussen et al.,[5] performed the first computations of the alpha decay fine structure for rotational nuclei using the coupled channel methods and later in 1958, the authors [6] developed an analytical method to estimate the alpha decay intensities to various rotational levels. On the basis of the nuclear shell model, in 1960, H. J. Mang [7] studied the decay rates for the ground state transitions of polonium isotopes and odd-even astatine isotopes and a good agreement with the experimental data was obtained. In 1976, J. O. Rasmussen et al., [8] applied new normalized and an older shell model alpha decay theory to predict decay rates of three 125 isotones of $^{209}$Po, $^{211}$Rn and $^{213}$Ra. By using simple cluster model, Buck et al [9, 10, 11] analyzed α-decay half-lives of favored transitions on even and odd mass nuclei. They could reproduce the experimental values with in a factor of two in the case of even-even nuclei and factor of three in the case of odd mass nuclei. To study the fine structure in α-decay of even-even neutron deficient nuclei in lead region, Richards et al [12] has used a modified two potential approach to the decay of quasi stationary state in conjunction with the particle-plus-rotor model. Mirea [13], using a theory, which was developed intending to describe cluster decay fine structure phenomena, based on Landau-Zener effect, explained α-decay fine structure of odd nuclei. Here fine structure in odd nuclei is explained by taking into account the radial and rotational coupling between the unpaired valence nucleon and the core of the decaying system. Delion et al [14] delivered a systematic microscopic description of α-decay to excited states in even-even spherical nuclei by taking the residual force as simple surface delta interaction. The low lying collective excitations are considered within the quasi particle random phase approximation. Calculations have been done exclusively on rotational and vibrational states with various potentials [15, 16, 17, 18]. Using stationary coupled channel approach [19] the authors reexamined the α-decay to $0^+$ and $2^+$ states by taking into accounts the deformation of nuclei. Using the Generalized Liquid Drop Model (GLDM) Wang et al [20, 21] predicted successfully the α-decay branching ratios to ground state rotational bands and excited $0^+$ states of even-even nuclei. Recently Ni et al [22] evaluated α-decay fine structure of even and odd nuclei by using Generalized Density Dependent Cluster Model (GDDCM). In this model the double folding nuclear potential is obtained by multipole expansion method. Most recently Ni et al [23, 24] subjected an exclusive study of α-transition to four different channels in even-even nuclei by taking axially deformed Woods-Saxon potential instead of multipole expansion method. By extending the semi empirical model to deformed nuclei, Denisov et al,.[25] proposed a model called unified model for alpha decay and alpha capture (UMADAC) and successfully evaluated alpha decay fine structure of even-even nuclei on a wide range. Within the framework of Coulomb and proximity potential model for deformed nuclei (CPPMDN), Santhosh et al [26, 27, 28] has done an elaborate study on α-decay fine structure of even-even nuclei and even-odd nuclei.

In this literature, using our recently proposed formalism - the Coulomb and proximity potential model for deformed nuclei (CPPMDN), we have done the fine structure study of odd-even nuclei in the range $83 \leq Z \leq 101$. We have shown that CPPMDN is successful in explaining alpha decay from ground and isomeric state [26], alpha fine structure of even-even [27] and even-odd nuclei [28], the present attempt is to establish CPPMDN as a unified theory of alpha-decay. The CPPMDN is the modified form of Coulomb and proximity potential model (CPPM) [29] by incorporating the quadrupole ($\beta_2$) and hexadecapole ($\beta_4$) deformations of parent and daughter nuclei. The external drifting potential barrier is obtained as the sum of deformed Coulomb potential, deformed two term proximity potential and centrifugal potential. Here the value of the parameters is kept the same as used in the study of even-even [27] and even-odd nuclei [28]. The importance of this one among the various other studies is that we could tackle the property, α-fine structure, which is purely microscopic, in terms of collective behavior by using a macroscopic model. We would like to mention that, this is the first extensive study on α-decay fine structure of odd-even nuclei, exclusively on both favored and unfavored, with a macroscopic approach.

The details of CPPMDN formalism is described in Sec. 2. In the following Sec.3 results and discussion are given and the entire work is concluded in Sec. 4.

## 2. The Coulomb and proximity potential model for deformed nuclei (CPPMDN)

In Coulomb and proximity potential model for deformed nuclei (CPPMDN), the potential energy barrier is taken as the sum of deformed Coulomb potential, deformed two term proximity potential and centrifugal potential for the touching configuration and for the separated fragments. For the pre-scission (overlap) region, simple power law interpolation as done by Shi and Swiatecki [30] is used. The inclusion of proximity potential reduces the height of the potential barrier, which closely agrees with the experimental result. We have shown that (see Table 1 of Ref [31]) the potential barrier computed for $^{14}$C and $^{24}$Ne radioactivity using Coulomb and proximity potential agrees with the experimental values obtained using the relation [32] given as

$$V(r) = 10.107 + 0.1021 Z_1 Z_2 - Q \qquad (1)$$

The proximity potential was first used by Shi and Swiatecki [30] in an empirical manner and has been quite extensively used over a decade by Gupta et al [33] in the Preformed Cluster Model (PCM). R K Puri et. al., [34, 35] has been using different versions of proximity potential for studying fusion cross section of different target-projectile combinations. In our model contribution of both internal and external part of the barrier is considered for the penetrability calculation. In present model assault frequency, ν is calculated for each parent-cluster combination which is associated with vibration energy. But Shi and Swiatecki [36] get ν empirically, unrealistic values $10^{22}$ for even A parent and $10^{20}$ for odd A parent.

The interacting potential barrier for two spherical nuclei is given by

$$V = \frac{Z_1 Z_2 e^2}{r} + V_p(z) + \frac{\hbar^2 \ell(\ell+1)}{2\mu r^2} \qquad \text{, for } z > 0 \qquad (2)$$

Here $Z_1$ and $Z_2$ are the atomic numbers of the daughter and emitted cluster, 'z' is the distance between the near surfaces of the fragments, 'r' is the distance between fragment centers, $\ell$ represents the angular momentum, $\mu$ the reduced mass, $V_P$ is the proximity potential given by Blocki et al [37] as

$$V_p(z) = 4\pi\gamma b \left[ \frac{C_1 C_2}{(C_1 + C_2)} \right] \Phi\left(\frac{z}{b}\right) \quad (3)$$

With the nuclear surface tension coefficient,

$$\gamma = 0.9517 [1 - 1.7826 (N - Z)^2 / A^2] \quad \text{MeV/fm}^2 \quad (4)$$

where N, Z and A represent neutron, proton and mass number of parent, $\Phi$ represent the universal the proximity potential [38] given as

$$\Phi(\varepsilon) = -4.41 e^{-\varepsilon / 0.7176}, \text{ for } \varepsilon \geq 1.9475 \quad (5)$$

$$\Phi(\varepsilon) = -1.7817 + 0.9270\varepsilon + 0.0169\varepsilon^2 - 0.05148\varepsilon^3 \text{ for } 0 \leq \varepsilon \leq 1.9475 \quad (6)$$

With $\varepsilon = z/b$, where the width (diffuseness) of the nuclear surface b $\approx$1 and Süsmann central radii $C_i$ of fragments related to sharp radii $R_i$ is

$$C_i = R_i - \left(\frac{b^2}{R_i}\right) \quad (7)$$

For $R_i$ we use semi empirical formula in terms of mass number $A_i$ as [37]

$$R_i = 1.28 A_i^{1/3} - 0.76 + 0.8 A_i^{-1/3} \quad (8)$$

The potential for the internal part (overlap region) of the barrier is given as

$$V = a_0 (L - L_0)^n \quad \text{for } z < 0 \quad (9)$$

where $L = z + 2C_1 + 2C_2$ and $L_0 = 2C$, the diameter of the parent nuclei. The constants $a_0$ and n are determined by the smooth matching of the two potentials at the touching point.

Using one dimensional WKB approximation, the barrier penetrability P is given as

$$P = \exp\left\{-\frac{2}{\hbar}\int_a^b \sqrt{2\mu(V-Q)}\,dz\right\} \tag{10}$$

Here the mass parameter is replaced by $\mu = mA_1A_2/A$, where m is the nucleon mass and $A_1$, $A_2$ are the mass numbers of daughter and emitted cluster respectively. The turning points "a" and "b" are determined from the equation, $V(a) = V(b) = Q$. The above integral can be evaluated numerically or analytically, and the half life time is given by

$$T_{1/2} = \left(\frac{\ln 2}{\lambda}\right) = \left(\frac{\ln 2}{\upsilon P}\right) \tag{11}$$

Where, $\upsilon = \left(\frac{\omega}{2\pi}\right) = \left(\frac{2E_v}{h}\right)$ represent the number of assaults on the barrier per second and $\lambda$ the decay constant. $E_v$, the empirical vibration energy is given as [39]

$$E_v = Q\left\{0.056 + 0.039\exp\left[\frac{(4-A_2)}{2.5}\right]\right\} \quad \text{for } A_2 \geq 4 \tag{12}$$

For alpha decay, $A_2 = 4$ and therefore the empirical vibration energy becomes,

$$E_v = 0.095Q \tag{13}$$

In the classical method, the $\alpha$ particle is assumed to move back and forth in the nucleus and the usual way of determining the assault frequency is through the expression given by $v = velocity/(2R)$, where R is the radius of the parent nuclei. But the alpha particle has wave properties; therefore a quantum mechanical treatment is more accurate. Thus, assuming that the alpha particle vibrates in a harmonic oscillator potential with a frequency $\omega$, which depends on the vibration energy $E_v$, we can identify this frequency as the assault frequency $v$ given in eqns. (11) – (13).

The Coulomb interaction between the two deformed and oriented nuclei taken from [40] with higher multipole deformation included [41, 42] is given as,

$$V_C = \frac{Z_1 Z_2 e^2}{r} + 3Z_1 Z_2 e^2 \sum_{\lambda, i=1,2} \frac{1}{2\lambda+1} \frac{R_{0i}^\lambda}{r^{\lambda+1}} Y_\lambda^{(0)}(\alpha_i) \left[ \beta_{\lambda i} + \frac{4}{7} \beta_{\lambda i}^2 Y_\lambda^{(0)}(\alpha_i) \delta_{\lambda,2} \right] \quad (14)$$

with

$$R_i(\alpha_i) = R_{0i} \left[ 1 + \sum_\lambda \beta_{\lambda i} Y_\lambda^0(\alpha_i) \right] \quad (15)$$

where $R_{0i} = 1.28 A_i^{1/3} - 0.76 + 0.8 A_i^{-1/3}$. Here $\alpha_i$ is the angle between the radius vector and symmetry axis of the $i^{th}$ nuclei (see Fig.1 of Ref [41]). Note that the quadrupole interaction term proportional to $\beta_{21}\beta_{22}$ is neglected because of its short range character.

Nuclear interaction [43, 44] can be taken into two variants: the proximity potential and the double folding potential. The latter is more effective in the description of interaction between two fragments. The proximity potential of Blocki et. Al., [37, 38] has one term based on the first approximation of the folding procedure, which describes the interaction between two pure spherically symmetric fragments. The two-term proximity potential of Baltz et. al., (equation (11) of [45]) includes the second component as the second approximation of the more accurate folding procedure. The authors have shown that the two-term proximity potential is in excellent agreement with the folding model for heavy ion reaction, not only in shape but also in absolute magnitude (see figure 3 of [45]). The two-term proximity potential for interaction between a deformed and spherical nucleus is given by Baltz et. al., [45] as

$$V_{P2}(R,\theta) = 2\pi \left[ \frac{R_1(\alpha) R_C}{R_1(\alpha) + R_C + S} \right]^{1/2} \left[ \frac{R_2(\alpha) R_C}{R_2(\alpha) + R_C + S} \right]^{1/2}$$

$$x\left[\left[\varepsilon_0(S)+\frac{R_1(\alpha)+R_C}{2R_1(\alpha)R_C}\varepsilon_1(S)\right]\left[\varepsilon_0(S)+\frac{R_2(\alpha)+R_C}{2R_2(\alpha)R_C}\varepsilon_1(S)\right]\right]^{1/2} \quad (16)$$

Here $R_1(\alpha)$ and $R_2(\alpha)$ are the principal radii of curvature of the daughter nuclei at the point where polar angle is $\alpha$, $S$ is the distance between the surfaces along the straight line connecting the fragments, $R_C$ is the radius of the spherical cluster, $\varepsilon_0(S)$ and $\varepsilon_1(S)$ are the one dimensional slab-on-slab function.

### 3. Results and discussion

In this study, the α-decay partial half life and branching ratio for each transition of odd-even nuclei in the range $83 \leq Z \leq 101$ are evaluated, by using the Coulomb and proximity potential model for deformed nuclei (CPPMDN). The entire calculation, that we have done, is tabulated in Table 1. In column one, different transitions are arranged by specifying the spin-parity of parent nucleus and that of different states of daughter nucleus. States having not well defined spin-parity values are mentioned in brackets. The state for which the spin-parity is not yet known is denoted by a question mark and in calculation they are treated as favored transitions. All these transitions are taken from Ref [46]. The energy of the emitted α-particle, Q - value is given in column two and is obtained using the following relation,

$$Q_i = Q_{gs \to gs} - E_i^* \quad (17)$$

where $Q_{gs \to gs}$ is the Q-value for ground state to ground state transition and $E^*$ is the excitation energy of daughter nucleus to the $i^{th}$ state. The ground state to ground state Q-value is obtained [25] by the expression,

$$Q_{gs \to gs} = \Delta M_p - (\Delta M_d + \Delta M_\alpha) + k(Z_p^\varepsilon - Z_d^\varepsilon) \quad (18)$$

where $\Delta M_p$, $\Delta M_d$ and $\Delta M_\alpha$ are correspondingly the mass excess of parent, daughter and α particle. The last term in eq. (18) is for incorporating the correction due to screening effect of electrons on ejecting alpha particle, where k=8.7 eV and ε = 2.517 for nuclei with Z ≥ 60 and k = 13.6 eV and ε = 2.408 for nuclei with Z ≤ 60. The mass excess is taken from Ref [47]; since these values are experimental the full shell effects can be included in this study. The value of angular momentum $\ell_\alpha$, which is used for calculation of centrifugal term, is arranged in column three. These are the minimum values of possible angular momenta; the α-particle can transfer any value of angular momentum which follows the equation (19),

$$\left|I_j - I_i\right| \leq \ell_\alpha \leq I_j + I_i \quad \text{and} \quad \frac{\pi_i}{\pi_j} = (-1)^{\ell_\alpha} \tag{19}$$

where $I_j$, $\pi_j$ and $I_i$, $\pi_i$ are the spin and parity of parent and daughter nuclei, respectively. For the evaluation of deformed Coulomb and nuclear potential, the experimental deformation values are taken from Ref [48], and where there is no experimental deformation values are available, theoretical values are taken from Ref [49].

The calculated partial half lives are given in column four. To get a comparison, in the next column the experimental partial half-life values are arranged. It can be evaluated from Ref [46], by using total α-half life and intensity to different states of daughter nucleus. At a glance, most calculated half-lives, irrespective of ground state to ground state or ground state to excited state, are in good agreement with the experimental one. Even though experimental half-lives span over a wide range of order, the calculated values are within an order of three. Transitions showing large deviations between calculated and experimental values are corresponding to less intense transitions and they have much longer half lives compared to other probable transitions. The comparison of calculated total half-life with the experimental value is shown in Fig.1. From this it is clear that, calculated values are in good agreement with the experimental results.

The calculated and experimental values of branching ratio are arranged in column six and seven respectively. It has been evaluated using the expressions in terms of width of α-decay Γ, given as

$$B_i = \frac{\Gamma(Q_i, \ell_i)}{\sum_n (Q_n, \ell_n)} \times 100\% \tag{20}$$

and

$$\Gamma(Q_i, \ell) = \hbar v \frac{1}{2} \int_0^\pi P(Q_i, \theta, \ell) \sin(\theta) d\theta \tag{21}$$

where v is the assault frequency and $P(Q_i, \theta, \ell)$ is the penetrability of the α particle in a direction θ from the symmetry axis, for axially symmetric deformed nuclei. From the comparison, one can see that, in most of the transitions, calculated branching ratios are close to the experimental values. Ground to ground transitions have more intensity and can be very effectively reproduced using our formulation. And our formulism is also successful to predict angular momentum hindered transitions. But in some elements, few favored transitions show a much deeper affinity to certain excited levels of daughter nuclei. This may be due to microscopic properties of nuclear structure, and hence structure hindrance may govern the prominent role of decay process. In Fig.2, α-decay from the ground state of $^{229}$Pa isotope to some levels of $^{225}$Ac is shown. The intensity to the various levels, spin-parity and energy of the states are specified. The hindrance factor (HF) is calculated by taking the ratio of experimental partial half-life with the calculated one. Transitions having low intensity correspond to high hindrance factor, or we can say such transitions are highly hindered.

The Fig.3 represents the plot for $\log_{10}(T_{1/2})$ versus the parent nuclei. The $\beta_2$ values for the corresponding parent nuclei are also shown in the graph. A systematic study on the g.s→g.s decay of the parent nuclei (eg. $^{215}$At, $^{225}$Pa, $^{227}$Pa, $^{229}$Pa, $^{235}$Np, $^{255}$Md, $^{257}$Md) and its decay products was done and this provided us a general trend that as the deformation values decreases, the calculated half lives (and also Hindrance Factor, HF) decreases. On examining the case of $^{235}$Np ($\beta_2$ = 0.215, $T_{1/2}$ = 9.74x10$^{12}$ s, HF = 9.014) and its decay product $^{231}$Pa ($\beta_2$ = 0.198, $T_{1/2}$ = 1.37x10$^{12}$ s, HF = 6.839), we observe that as the $\beta_2$ decreases, the HF decreases or vice versa. i.e as the $\beta_2$ value increases the HF increases which indicates structural hindrance. The same trend is observed for the parents $^{225}$Pa, $^{227}$Pa and $^{229}$Pa which are clear from the graph. But in the case of parents $^{215}$At, $^{255}$Md and $^{257}$Md the trend seen is just the opposite. This reverse trend is due to the presence of the neutron shell closure of the daughters $^{207}$Tl (N = 126), $^{247}$Bk (N ≈ 152) and $^{249}$Bk (N = 152) respectively in the corresponding decay chain. This reveals the fact that in general as the deformation values decreases, the decay is less hindered. In the case of decay chain of $^{229}$Pa it can be seen that the $\log_{10}(T_{1/2})$ decreases with decrease in $\beta_2$ values up to the daughter $^{213}$Bi and thereafter increases with decrease in $\beta_2$ value. This is also due to the near neutron magicity (N = 128) of the daughter nuclei $^{209}$Tl present in the decay chain. An examination of the graph also shows a similar behaviour for the ground state to excited state decays as that of g.s→g.s decays. Thus it is clear that our model is able to predict conclusively the transitions to the excited states also. We would also like to mention that for computing Q values, the experimental mass excess is taken from Ref [47]. So full shell effects are incorporated in our model that comes through experimental Q values. Thus our study reveals that the differences in the parent and daughter surfaces or the changes in the deformation parameters as well as the shell structure of the parent and daughter nuclei, influences the alpha decay probability. This study agrees well with the findings of J. O. Rasmussen and collaborators [50].

In our study, for some nuclei we have observed large differences between experimental and predicted rates. It is found that the g.s→g.s decay of $^{211}$Bi ($\beta_2$ = -0.018) and $^{213}$Bi ($\beta_2$ = -0.018) have high HF values namely 13.05 and 5.97 respectively. The daughter nuclei of these decays are

$^{207}$Tl ($\beta_2$ = -0.008) and $^{209}$Tl ($\beta_2$ = -0.008) respectively. In both these cases, the deformation values increase from -0.018 to -0.008. From this it is clear that as the $\beta_2$ value increases, the HF also increases. In both the cases it should be also noted that the high HF values are also due to the presence of the proton and neutron magicity of the daughters $^{207}$Tl (Z ≈ 82, N = 126) and near proton and neutron magicity of $^{209}$Tl (Z ≈ 82, N ≈ 126) respectively. We would also like to highlight the fact that the highest HF is obtained for that decay leading to near doubly magic $^{207}$Tl (Z ≈ 82, N = 126). These facts reveal that the large differences between the experimental and predicted decay rates depend on the deformation values and the presence of shell closure.

Fig.4 represents the variation of $\log_{10}$ ($T_{1/2}$) for different angular momentum values for the nuclei $^{221}$Fr and its decay product $^{217}$At separately. It is clear that as the angular momentum increases, the centrifugal barrier plays a prominent role and the decay becomes angular momentum hindered. Now when we compare the half life of $^{217}$At ($\beta_2$ = 0.039) with that of $^{221}$Fr ($\beta_2$ = 0.120), it can be seen that the half life value increases as the $\ell_\alpha$ value and $\beta_2$ increases. The HF values for the corresponding $\ell_\alpha$ values also show an increment. These facts reveal the interplay of angular hindrance and nuclear structure hindrance.

Before concluding this study on odd-even nuclei, to show the effective strength of our formulation, we have evaluated the standard deviation value of both half-life and branching ratio using the equation,

$$\sigma = \sqrt{\frac{1}{(n-1)}\sum_{i=1}^{n}\left[\log\left(\frac{T_{1/2}^{cal.}}{T_{1/2}^{exp.}}\right)\right]^2} \qquad (22)$$

The obtained standard deviation of half-life of all transitions is 1.08 and that for branching ratio is 1.21. In Fig.5. the comparison of calculated partial half-life values with experimental data for entire transitions has been done. From this it is clear that, for most of the transitions, the deviation of calculated value with the corresponding experimental value lie within the order two.

### 4. Summary

In summary, we have evaluated alpha decay half lives of odd-even nuclei on a wide range using the recently proposed CPPMDN formalism. Here we restrict the study on decay of nuclei from ground state of parent nuclei to ground and excited states of daughter nucleus. The α-decay intensities to various states are evaluated and it is compared with the experimental values. The order of standard deviation value is nearly unity for half-life and exact one for branching ratio. So as a conclusion we can say that the CPPMDN is apt for macroscopic study of alpha fine structure in odd-even nuclei. The present study reveals that CPPMDN is a unified theory which is successful in explaining alpha decay from ground and isomeric state; for alpha fine structure of even-even, even-odd and odd-even nuclei.

Table 1. Comparison of computed half lives and branching ratios of nuclei in the range $83 \leq Z \leq 101$ with the corresponding experimental values. Here Q-values are in MeV and half-lives in seconds.

| Transitions | Q | $\ell_{min}$ | $T_{cal.}$ | $T_{exp.}$ | $B_{cal.}$ | $B_{exp}$ |
|---|---|---|---|---|---|---|
| $^{211}Bi \rightarrow ^{207}Tl$ | | | | | | |
| $9/2^- \rightarrow 1/2^+$ | 6.7861 | 5 | $1.18 \times 10^1$ | $1.54 \times 10^2$ | 89.10 | 83.77 |
| $9/2^- \rightarrow 3/2^+$ | 6.4350 | 3 | $9.64 \times 10^1$ | $7.93 \times 10^2$ | 10.90 | 16.23 |
| | | | | | | |
| $^{213}Bi \rightarrow ^{209}Tl$ | | | | | | |
| $9/2^- \rightarrow (1/2^+)$ | 6.0171 | 5 | $2.36 \times 10^4$ | $1.41 \times 10^5$ | 92.30 | 92.81 |
| $9/2^- \rightarrow (3/2^+)$ | 5.6933 | 3 | $2.83 \times 10^5$ | $1.82 \times 10^6$ | 7.70 | 7.19 |
| | | | | | | |
| $^{211}At \rightarrow ^{207}Bi$ | | | | | | |
| $9/2^- \rightarrow 9/2^-$ | 6.0187 | 0 | $7.78 \times 10^4$ | $6.21 \times 10^4$ | 99.97 | 99.99 |
| $9/2^- \rightarrow 11/2^-$ | 5.3491 | 2 | $4.08 \times 10^8$ | $7.21 \times 10^8$ | $1.91 \times 10^{-2}$ | $8.61 \times 10^{-3}$ |
| $9/2^- \rightarrow 7/2^-$ | 5.2760 | 2 | $1.09 \times 10^9$ | $2.60 \times 10^9$ | $7.14 \times 10^{-3}$ | $2.39 \times 10^{-3}$ |
| $9/2^- \rightarrow 9/2^-$ | 5.1327 | 0 | $5.42 \times 10^9$ | $6.21 \times 10^9$ | $1.44 \times 10^{-3}$ | $1.00 \times 10^{-3}$ |
| | | | | | | |
| $^{215}At \rightarrow ^{211}Bi$ | | | | | | |
| $9/2^- \rightarrow 9/2^-$ | 8.2144 | 0 | $1.21 \times 10^{-4}$ | $1.00 \times 10^{-4}$ | 96.86 | 99.95 |
| $9/2^- \rightarrow 7/2^-$ | 7.8064 | 2 | $3.73 \times 10^{-3}$ | $2.00 \times 10^{-1}$ | 3.14 | 0.05 |
| | | | | | | |
| $^{217}At \rightarrow ^{213}Bi$ | | | | | | |
| $9/2^- \rightarrow 9/2^-$ | 7.2384 | 0 | $1.54 \times 10^{-1}$ | $3.23 \times 10^{-2}$ | 94.96 | 99.94 |
| $9/2^- \rightarrow 7/2^-$ | 6.9805 | 2 | 3.12 | 89.72 | 4.69 | $3.60 \times 10^{-2}$ |
| $9/2^- \rightarrow (5/2,7/2,9/2)^-$ | 6.6452 | 0 | 47.23 | 153.81 | 0.31 | $2.10 \times 10^{-2}$ |
| $9/2^- \rightarrow (5/2,13/2)^-$ | 6.4795 | 2 | $3.60 \times 10^2$ | $6.46 \times 10^2$ | $4.06 \times 10^{-2}$ | $5.00 \times 10^{-3}$ |
| $9/2^- \rightarrow ?$ | 6.1884 | 0 | $3.65 \times 10^3$ | $3.23 \times 10^3$ | $4.01 \times 10^{-3}$ | $9.99 \times 10^{-4}$ |

| Transitions | Q | $\ell_{min}$ | $T_{cal.}$ | $T_{exp.}$ | $B_{cal.}$ | $B_{exp}$ |
|---|---|---|---|---|---|---|
| $^{221}$Fr → $^{217}$At | | | | | | |
| 5/2⁻→9/2⁻ | 6.4948 | 2 | $2.25 \times 10^2$ | $3.53 \times 10^2$ | 82.73 | 82.71 |
| 5/2⁻→(7/2⁻) | 6.1125 | 2 | $1.18 \times 10^5$ | $3.68 \times 10^5$ | $1.58 \times 10^{-1}$ | $7.93 \times 10^{-2}$ |
| 5/2⁻→7/2⁻ | 6.3945 | 2 | $1.99 \times 10^3$ | $2.19 \times 10^4$ | 9.35 | 1.33 |
| 5/2⁻→5/2⁻ | 6.2767 | 0 | $4.83 \times 10^3$ | $1.95 \times 10^3$ | 3.85 | 14.97 |
| 5/2⁻→3/2⁻ | 6.2227 | 2 | $1.24 \times 10^4$ | $1.96 \times 10^5$ | 1.50 | 0.15 |
| 5/2⁻→? | 6.1845 | 0 | $1.31 \times 10^4$ | $9.80 \times 10^6$ | 1.42 | $2.98 \times 10^{-3}$ |
| 5/2⁻→(3/2)⁻ | 6.1266 | 2 | $3.54 \times 10^4$ | $6.00 \times 10^4$ | 0.53 | 0.49 |
| 5/2⁻→13/2⁻ | 6.0842 | 4 | $1.26 \times 10^5$ | $1.73 \times 10^5$ | 0.15 | 0.17 |
| 5/2⁻→(5/2,7/2,9/2)⁻ | 6.0704 | 0 | $1.27 \times 10^5$ | $9.80 \times 10^5$ | 0.15 | $2.98 \times 10^{-2}$ |
| 5/2⁻→(9/2⁺) | 5.9573 | 3 | $1.04 \times 10^6$ | $7.35 \times 10^6$ | $1.79 \times 10^{-2}$ | $3.97 \times 10^{-3}$ |
| 5/2⁻→(7/2,9/2) | 5.9263 | 0 | $2.43 \times 10^5$ | $5.88 \times 10^6$ | $7.66 \times 10^{-2}$ | $4.97 \times 10^{-3}$ |
| 5/2⁻→(7/2)⁻ | 5.9173 | 2 | $3.84 \times 10^5$ | $4.90 \times 10^5$ | $4.85 \times 10^{-2}$ | $5.96 \times 10^{-2}$ |
| 5/2⁻→? | 5.8428 | 0 | $1.82 \times 10^6$ | $2.94 \times 10^7$ | $1.02 \times 10^{-2}$ | $9.93 \times 10^{-4}$ |
| 5/2⁻→? | 5.8304 | 0 | $2.11 \times 10^6$ | $1.47 \times 10^7$ | $8.82 \times 10^{-3}$ | $1.99 \times 10^{-3}$ |
| 5/2⁻→? | 5.6855 | 0 | $1.25 \times 10^7$ | $3.27 \times 10^7$ | $1.49 \times 10^{-3}$ | $8.93 \times 10^{-4}$ |
| | | | | | | |
| $^{215}$Ac → $^{211}$Fr | | | | | | |
| 9/2⁻→9/2⁻ | 7.7841 | 0 | 0.12 | 0.17 | 98.70 | 99.21 |
| 9/2⁻→? | 7..3851 | 0 | 11.33 | 36.96 | 1.05 | 0.46 |
| 9/2⁻→(11/2⁻) | 7.2018 | 2 | 85.27 | 85.00 | 0.14 | 0.20 |
| 9/2⁻→(13/2⁻) | 7.1301 | 2 | 103.76 | 121.43 | 0.11 | 0.14 |
| | | | | | | |
| $^{221}$Ac → $^{217}$Fr | | | | | | |
| 9/2⁻→9/2⁻ | 7.8191 | 0 | $7.18 \times 10^{-2}$ | $7.65 \times 10^{-2}$ | 69.67 | 67.88 |
| 9/2⁻→? | 7.6101 | 0 | 0.24 | 0.25 | 20.84 | 20.77 |
| 9/2⁻→? | 7.5461 | 0 | 0.67 | 0.58 | 7.47 | 8.95 |
| 9/2⁻→? | 7.3351 | 0 | 2.48 | 2.17 | 2.02 | 2.39 |

| Transitions | Q | $\ell_{min}$ | $T_{cal.}$ | $T_{exp.}$ | $B_{cal.}$ | $B_{exp}$ |
|---|---|---|---|---|---|---|
| $^{223}$Ac → $^{219}$Fr | | | | | | |
| (5/2⁻)→9/2⁻ | 6.8221 | 2 | 6.62x10² | 4.06x10² | 20.12 | 13.05 |
| (5/2⁻)→5/2⁻ | 6.8071 | 0 | 5.12x10² | 2.86x10² | 26.02 | 18.52 |
| (5/2⁻)→(3/2⁻) | 6.7660 | 2 | 1.15x10³ | 2.52x10² | 11.58 | 21.02 |
| (5/2⁻)→(1/2⁻) | 6.7411 | 2 | 1.47x10³ | 4.20x10² | 9.06 | 12.61 |
| (5/2⁻)→(7/2)⁻ | 6.7235 | 2 | 1.75x10³ | 9.26x10² | 7.61 | 5.72 |
| (5/2⁻)→(5/2)⁻ | 6.6877 | 0 | 1.68x10³ | 4.06x10³ | 7.93 | 1.30 |
| (5/2⁻)→(3/2⁻) | 6.6823 | 2 | 2.65x10³ | 2.10x10² | 5.03 | 25.22 |
| (5/2⁻)→(7/2)⁺ | 6.6308 | 3 | 6.61x10³ | 4.06x10³ | 2.02 | 1.30 |
| (5/2⁻)→(3/2⁺) | 6.6117 | 1 | 4.15x10³ | 2.10x10⁵ | 3.21 | 2.52x10⁻² |
| (5/2⁻)→(11/2⁺) | 6.6061 | 3 | 8.50x10³ | 6.30x10⁴ | 1.57 | 8.41x10⁻² |
| (5/2⁻)→(7/2⁻) | 6.5529 | 2 | 9.91x10³ | 9.77x10⁴ | 1.34 | 5.42x10⁻² |
| (5/2⁻)→(9/2⁻) | 6.5166 | 2 | 1.45x10⁴ | 5.73x10⁴ | 9.19x10⁻¹ | 9.24x10⁻² |
| (5/2⁻)→(11/2⁻) | 6.4886 | 4 | 4.79x10⁴ | 9.00x10⁴ | 2.78x10⁻¹ | 5.89x10⁻² |
| (5/2⁻)→(5/2⁺) | 6.4818 | 1 | 1.60x10⁴ | 4.20x10⁴ | 8.33x10⁻¹ | 1.26x10⁻¹ |
| (5/2⁻)→(7/2⁺) | 6.4497 | 1 | 2.24x10⁴ | 2.68x10⁴ | 5.95x10⁻¹ | 1.98x10⁻¹ |
| (5/2⁻)→(7/2) | 6.4473 | 1 | 2.20x10⁴ | 2.14x10⁵ | 6.06x10⁻¹ | 2.48x10⁻² |
| (5/2⁻)→(5/2⁺) | 6.4378 | 1 | 2.54x10⁴ | 2.52x10⁵ | 5.24x10⁻¹ | 2.10x10⁻² |
| (5/2⁻)→(9/2) | 6.3901 | 2 | 8.17x10⁴ | 1.41x10⁵ | 1.63x10⁻¹ | 3.76x10⁻² |
| (5/2⁻)→(9/2⁺) | 6.3599 | 3 | 1.13x10⁵ | 4.20x10⁵ | 1.18x10⁻¹ | 1.26x10⁻² |
| (5/2⁻)→(5/2⁻) | 6.3318 | 0 | 6.98x10⁴ | 1.34x10⁴ | 1.91x10⁻¹ | 3.95x10⁻¹ |
| (5/2⁻)→(9/2⁺) | 6.3156 | 3 | 1.83x10⁵ | 2.55x10⁵ | 7.28x10⁻² | 2.08x10⁻² |
| (5/2⁻)→(11/2⁺) | 6.2921 | 3 | 2.37x10⁵ | 4.20x10⁵ | 5.62x10⁻² | 1.26x10⁻² |
| (5/2⁻)→(7/2⁻) | 6.2883 | 2 | 1.67x10⁵ | 1.05x10⁵ | 7.98x10⁻² | 5.04x10⁻² |
| (5/2⁻)→(9/2⁻) | 6.2331 | 2 | 3.08x10⁵ | 4.24x10⁵ | 4.33x10⁻² | 1.25x10⁻² |
| (5/2⁻)→(11/2⁻) | 6.1721 | 4 | 1.50x10⁶ | 1.27x10⁶ | 8.88x10⁻³ | 4.17x10⁻³ |
| (5/2⁻)→(5/2⁺) | 6.1166 | 1 | 8.85x10⁵ | 4.24x10⁵ | 1.51x10⁻² | 1.25x10⁻² |
| (5/2⁻)→(7/2⁺) | 6.0441 | 1 | 2.05x10⁶ | 1.27x10⁶ | 6.50x10⁻³ | 4.17x10⁻³ |

| Transitions | Q | $\ell_{min}$ | $T_{cal.}$ | $T_{exp.}$ | $B_{cal.}$ | $B_{exp}$ |
|---|---|---|---|---|---|---|
| $^{225}$Ac → $^{221}$Fr | | | | | | |
| (3/2⁻)→5/2⁻ | 5.9741 | 2 | 5.63x10⁶ | 1.70x10⁶ | 19.41 | 50.60 |
| (3/2⁻)→(1/2)⁻ | 5.9481 | 2 | 7.67x10⁶ | 2.88x10⁸ | 14.25 | 0.30 |
| (3/2⁻)→(9/2)⁻ | 5.9356 | 4 | 2.16x10⁷ | 1.00x10⁷ | 5.06 | 8.60 |
| (3/2⁻)→(3/2)⁻ | 5.9374 | 0 | 5.90x10⁶ | 4.77x10⁶ | 18.53 | 18.03 |
| (3/2⁻)→(3/2)⁻ | 5.8745 | 0 | 1.26x10⁷ | 1.08x10⁷ | 8.67 | 7.97 |
| (3/2⁻)→(1/2)⁺ | 5.8742 | 1 | 1.27x10⁷ | 6.55x10⁷ | 8.61 | 1.31 |
| (3/2⁻)→(5/2)⁻ | 5.8732 | 2 | 1.89x10⁷ | 9.93x10⁷ | 5.78 | 0.87 |
| (3/2⁻)→(7/2)⁻ | 5.8657 | 2 | 2.08x10⁷ | 2.79x10⁷ | 5.25 | 3.08 |
| (3/2⁻)→(1/2)⁺ | 5.8282 | 1 | 2.54x10⁷ | 3.76x10⁹ | 4.30 | 0.02 |
| (3/2⁻)→(7/2)⁺ | 5.8240 | 3 | 5.07x10⁷ | 6.65x10⁷ | 2.16 | 1.29 |
| (3/2⁻)→(7/2)⁻ | 5.7783 | 2 | 6.10x10⁷ | 1.96x10⁷ | 1.79 | 4.39 |
| (3/2⁻)→(3/2)⁺ | 5.7494 | 1 | 6.77x10⁷ | 7.85x10⁷ | 1.61 | 1.10 |
| (3/2⁻)→(5/2)⁺ | 5.7396 | 1 | 7.66x10⁷ | 2.16x10⁹ | 1.43 | 0.04 |
| (3/2⁻)→(5/2)⁺ | 5.7205 | 1 | 9.75x10⁷ | 7.20x10⁷ | 1.12 | 1.19 |
| (3/2⁻)→(7/2⁻,9/2⁻) | 5.7014 | 2 | 1.61x10⁸ | 2.54x10⁸ | 0.68 | 0.34 |
| (3/2⁻)→(7/2)⁺ | 5.6949 | 3 | 2.56x10⁸ | 8.64x10⁸ | 4.27x10⁻¹ | 9.96x10⁻² |
| (3/2⁻)→(9/2)⁻ | 5.6860 | 4 | 4.70x10⁸ | 2.88x10⁹ | 2.33 x10⁻¹ | 2.99x10⁻² |
| (3/2⁻)→(9/2)⁺ | 5.6794 | 3 | 3.12x10⁸ | 2.76x10⁹ | 3.50x10⁻¹ | 3.12x10⁻² |
| (3/2⁻)→(7/2,5/2)⁺ | 5.5808 | 3 | 1.12x10⁹ | 6.17x10⁸ | 9.76x10⁻² | 1.39x10⁻¹ |
| (3/2⁻)→(7/2⁻) | 5.5735 | 2 | 8.43x10⁸ | 1.23x10⁹ | 1.30x10⁻¹ | 6.99x10⁻² |
| (3/2⁻)→(5/2⁺) | 5.4563 | 1 | 3.13x10⁹ | 1.23x10⁹ | 3.49x10⁻² | 6.99x10⁻² |
| (3/2⁻)→(3/2)⁻ | 5.4220 | 0 | 4.40x10⁹ | 3.76x10⁸ | 2.48 x10⁻² | 2.29x10⁻¹ |
| (3/2⁻)→(7/2⁺,5/2⁺) | 5.4033 | 3 | 1.22x10¹⁰ | 6.17x10⁸ | 8.96x10⁻³ | 1.39x10⁻¹ |
| (3/2⁻)→(5/2⁻) | 5.3718 | 2 | 1.29x10¹⁰ | 2.88x10¹⁰ | 8.47x10⁻³ | 2.99x10⁻³ |
| (3/2⁻)→(5/2⁺) | 5.3434 | 1 | 1.49x10¹⁰ | 2.88x10⁹ | 7.34x10⁻³ | 2.99x10⁻² |
| (3/2⁻)→? | 5.3364 | 0 | 1.44x10¹⁰ | 4.32x10¹⁰ | 7.59x10⁻³ | 1.99x10⁻³ |
| (3/2⁻)→? | 5.2621 | 0 | 4.15x10¹⁰ | 4.32x10¹⁰ | 2.63x10⁻³ | 1.99x10⁻³ |

| Transitions | Q | $\ell_{min}$ | $T_{cal.}$ | $T_{exp.}$ | $B_{cal.}$ | $B_{exp}$ |
|---|---|---|---|---|---|---|
| $(3/2^-)\rightarrow?$ | 5.2599 | 0 | $4.28\times10^{10}$ | $4.32\times10^{10}$ | $2.55\times10^{-3}$ | $1.99\times10^{-3}$ |
| $(3/2^-)\rightarrow?$ | 5.2249 | 0 | $7.09\times10^{10}$ | $1.44\times10^{10}$ | $1.54\times10^{-3}$ | $5.97\times10^{-3}$ |
| $(3/2^-)\rightarrow?$ | 5.1939 | 0 | $1.11\times10^{11}$ | $2.88\times10^{10}$ | $9.85\times10^{-4}$ | $2.99\times10^{-3}$ |
| $(3/2^-)\rightarrow?$ | 5.1499 | 0 | $2.13\times10^{11}$ | $1.73\times10^{10}$ | $5.13\times10^{-4}$ | $4.97\times10^{-3}$ |
| $^{227}Ac \rightarrow {}^{223}Fr$ | | | | | | |
| $3/2^-\rightarrow3/2^{(-)}$ | 5.0812 | 0 | $5.54\times10^{11}$ | $1.04\times10^{11}$ | 38.48 | 46.32 |
| $3/2^-\rightarrow(5/2^-)$ | 5.0683 | 2 | $9.84\times10^{11}$ | $1.26\times10^{11}$ | 21.66 | 38.23 |
| $3/2^-\rightarrow1/2^{(-)}$ | 5.0262 | 2 | $1.87\times10^{12}$ | $4.53\times10^{13}$ | 11.40 | 0.11 |
| $3/2^-\rightarrow(7/2^-)$ | 4.9991 | 2 | $2.84\times10^{12}$ | $7.90\times10^{11}$ | 7.51 | 6.10 |
| $3/2^-\rightarrow(3/2^-)$ | 4.9816 | 0 | $2.56\times10^{12}$ | $8.28\times10^{11}$ | 8.33 | 5.82 |
| $3/2^-\rightarrow(3/2^+)$ | 4.9467 | 1 | $5.01\times10^{12}$ | $7.11\times10^{13}$ | 4.26 | 0.07 |
| $3/2^-\rightarrow(3/2^+)$ | 4.9207 | 1 | $7.55\times10^{12}$ | $4.91\times10^{12}$ | 2.82 | 0.98 |
| $3/2^-\rightarrow(5/2^+)$ | 4.9091 | 1 | $9.08\times10^{12}$ | $6.22\times10^{13}$ | 2.35 | 0.08 |
| $3/2^-\rightarrow(7/2^-)$ | 4.8921 | 2 | $1.53\times10^{13}$ | $2.75\times10^{12}$ | 1.39 | 1.75 |
| $3/2^-\rightarrow(7/2^+)$ | 4.8616 | 3 | $3.60\times10^{13}$ | $5.53\times10^{13}$ | $5.92\times10^{-1}$ | $8.71\times10^{-2}$ |
| $3/2^-\rightarrow(5/2)$ | 4.8386 | 0 | $2.49\times10^{13}$ | $1.15\times10^{13}$ | $8.56\times10^{-1}$ | $4.19\times10^{-1}$ |
| $3/2^-\rightarrow?$ | 4.7157 | 0 | $1.91\times10^{14}$ | $2.49\times10^{14}$ | $1.12\times10^{-1}$ | $1.93\times10^{-2}$ |
| $3/2^-\rightarrow?$ | 4.7102 | 0 | $2.09\times10^{14}$ | $4.98\times10^{14}$ | $1.02\times10^{-1}$ | $9.67\times10^{-3}$ |
| $3/2^-\rightarrow?$ | 4.7022 | 0 | $2.40\times10^{14}$ | $1.68\times10^{15}$ | $8.88\times10^{-2}$ | $2.87\times10^{-3}$ |
| $3/2^-\rightarrow?$ | 4.6322 | 0 | $7.96\times10^{14}$ | $1.68\times10^{15}$ | $2.68\times10^{-2}$ | $2.87\times10^{-3}$ |
| $3/2^-\rightarrow?$ | 4.5782 | 0 | $2.05\times10^{15}$ | $1.68\times10^{15}$ | $1.04\times10^{-2}$ | $2.87\times10^{-3}$ |
| $3/2^-\rightarrow3/2^-$ | 4.5660 | 0 | $2.54\times10^{15}$ | $9.96\times10^{14}$ | $8.39\times10^{-3}$ | $4.84\times10^{-3}$ |
| $3/2^-\rightarrow(5/2^+)$ | 4.5405 | 1 | $4.54\times10^{15}$ | $8.30\times10^{14}$ | $4.70\times10^{-3}$ | $5.80\times10^{-3}$ |
| $3/2^-\rightarrow(5/2^-)$ | 4.4802 | 2 | $1.72\times10^{16}$ | $1.68\times10^{15}$ | $1.24\times10^{-3}$ | $2.87\times10^{-3}$ |
| $^{217}Pa \rightarrow {}^{213}Ac$ | | | | | | |
| $9/2^-\rightarrow9/2^-$ | 8.7355 | 0 | $2.83\times10^{-3}$ | $3.64\times10^{-3}$ | 94.69 | 99.00 |

| Transitions | Q | $\ell_{min}$ | $T_{cal.}$ | $T_{exp.}$ | $B_{cal.}$ | $B_{exp}$ |
|---|---|---|---|---|---|---|
| 9/2⁻→? | 8.2675 | 0 | 8.17x10⁻² | 9.00x10⁻¹ | 3.28 | 0.40 |
| 9/2⁻→? | 8.1195 | 0 | 0.25 | 1.20 | 1.07 | 0.30 |
| 9/2⁻→? | 8.1015 | 0 | 0.28 | 1.20 | 0.96 | 0.30 |
| ²²⁵Pa → ²²¹Ac | | | | | | |
| 5/2⁻→(3/2⁻) | 7.4355 | 2 | 1.44 | 2.43 | 52.79 | 51.98 |
| 5/2⁻→? | 7.3835 | 0 | 1.61 | 2.63 | 47.21 | 48.02 |
| ²²⁷Pa → ²²³Ac | | | | | | |
| (5/2⁻)→(5/2⁻) | 6.6215 | 0 | 2.79x10⁴ | 5.38x10³ | 29.77 | 50.90 |
| (5/2⁻)→(7/2⁻) | 6.5791 | 2 | 6.45x10⁴ | 2.32x10⁴ | 12.88 | 11.80 |
| (5/2⁻)→(5/2⁻) | 6.5708 | 0 | 4.75x10⁴ | 1.80x10⁴ | 17.49 | 15.21 |
| (5/2⁻)→(5/2⁺) | 6.5569 | 1 | 6.28x10⁴ | 2.87x10⁴ | 13.23 | 9.54 |
| (5/2⁻)→(9/2⁻) | 6.5308 | 2 | 1.07x10⁵ | 1.04x10⁵ | 7.76 | 2.63 |
| (5/2⁻)→(7/2⁺) | 6.5114 | 1 | 1.02x10⁵ | 3.43x10⁴ | 8.14 | 7.98 |
| (5/2⁻)→(7/2⁺) | 6.4908 | 1 | 1.27x10⁵ | 3.86x10⁵ | 6.54 | 0.71 |
| (5/2⁻)→(11/2⁻) | 6.4805 | 4 | 4.44x10⁵ | 6.76x10⁵ | 1.87 | 0.41 |
| (5/2⁻)→(9/2⁺) | 6.4540 | 3 | 3.59x10⁵ | 3.38x10⁵ | 2.31 | 0.81 |
| ²²⁹Pa → ²²⁵Ac | | | | | | |
| (5/2⁺)→(3/2⁻) | 5.8755 | 1 | 1.45x10⁸ | 1.80x10¹⁰ | 24.62 | 0.15 |
| (5/2⁺)→(5/2⁻) | 5.8456 | 1 | 2.11x10⁸ | 2.59x10¹⁰ | 16.92 | 0.11 |
| (5/2⁺)→(3/2⁺) | 5.8354 | 2 | 3.08x10⁸ | 1.35x10¹⁰ | 11.59 | 0.20 |
| (5/2⁺)→(5/2⁺) | 5.8108 | 0 | 2.87x10⁸ | 1.46x10⁸ | 12.44 | 18.78 |
| (5/2⁺)→(7/2⁻) | 5.7984 | 1 | 3.82x10⁸ | 2.59x10¹⁰ | 9.35 | 0.11 |
| (5/2⁺)→(7/2⁺) | 5.7704 | 2 | 7.02x10⁸ | 2.76x10⁸ | 5.09 | 9.93 |
| (5/2⁺)→(5/2⁻) | 5.7547 | 1 | 6.68x10⁸ | 2.03x10⁸ | 5.34 | 13.51 |
| (5/2⁺)→(9/2⁺) | 5.7305 | 2 | 1.17x10⁹ | 5.86x10⁸ | 3.05 | 4.68 |

| Transitions | Q | $\ell_{min}$ | $T_{cal.}$ | $T_{exp.}$ | $B_{cal.}$ | $B_{exp}$ |
|---|---|---|---|---|---|---|
| $(5/2^+) \rightarrow (5/2^+)$ | 5.7198 | 0 | $9.22 \times 10^8$ | $7.49 \times 10^7$ | 3.87 | 36.61 |
| $(5/2^+) \rightarrow (7/2^-)$ | 5.7047 | 1 | $1.27 \times 10^9$ | $6.93 \times 10^8$ | 2.81 | 3.96 |
| $(5/2^+) \rightarrow (7/2^+)$ | 5.6756 | 2 | $2.40 \times 10^9$ | $3.16 \times 10^8$ | 1.49 | 8.68 |
| $(5/2^+) \rightarrow ?$ | 5.6545 | 0 | $2.17 \times 10^9$ | $4.47 \times 10^9$ | 1.65 | 0.61 |
| $(5/2^+) \rightarrow (9/2^-)$ | 5.6400 | 3 | $5.54 \times 10^9$ | $3.81 \times 10^9$ | 0.64 | 0.72 |
| $(5/2^+) \rightarrow (9/2^+)$ | 5.6185 | 2 | $5.09 \times 10^9$ | $1.58 \times 10^9$ | 0.70 | 1.74 |
| $(5/2^+) \rightarrow (11/2^-)$ | 5.5575 | 3 | $1.67 \times 10^{10}$ | $3.81 \times 10^{10}$ | 0.21 | 0.07 |
| $(5/2^+) \rightarrow (11/2^+)$ | 5.5485 | 4 | $3.05 \times 10^{10}$ | $2.70 \times 10^{10}$ | 0.12 | 0.10 |
| $(5/2^+) \rightarrow ?$ | 5.4545 | | $3.25 \times 10^{10}$ | $5.40 \times 10^{10}$ | 0.11 | 0.05 |
| $^{231}Pa \rightarrow {}^{227}Ac$ | | | | | | |
| $3/2^- \rightarrow 3/2^-$ | 5.1903 | 0 | $1.37 \times 10^{12}$ | $9.37 \times 10^{12}$ | 32.55 | 11.11 |
| $3/2^- \rightarrow 3/2^+$ | 5.1629 | 1 | $2.34 \times 10^{12}$ | $4.12 \times 10^{13}$ | 19.06 | 2.53 |
| $3/2^- \rightarrow 5/2^-$ | 5.1603 | 2 | $3.11 \times 10^{12}$ | $5.15 \times 10^{12}$ | 14.34 | 20.21 |
| $3/2^- \rightarrow 5/2^+$ | 5.1439 | 1 | $3.12 \times 10^{12}$ | $4.06 \times 10^{12}$ | 14.29 | 25.63 |
| $3/2^- \rightarrow (7/2)^-$ | 5.1162 | 2 | $6.08 \times 10^{12}$ | $7.36 \times 10^{13}$ | 7.33 | 1.41 |
| $3/2^- \rightarrow (7/2)^+$ | 5.1057 | 3 | $1.03 \times 10^{13}$ | $2.58 \times 10^{14}$ | 4.33 | 0.40 |
| $3/2^- \rightarrow (9/2)^+$ | 5.0804 | 3 | $1.51 \times 10^{13}$ | $4.52 \times 10^{12}$ | 2.95 | 23.02 |
| $3/2^- \rightarrow (9/2)^-$ | 5.0634 | 4 | $3.16 \times 10^{13}$ | $3.44 \times 10^{13}$ | 1.41 | 3.03 |
| $3/2^- \rightarrow ?$ | 5.0303 | 0 | $1.59 \times 10^{13}$ | $5.15 \times 10^{16}$ | 2.80 | $2.02 \times 10^{-3}$ |
| $3/2^- \rightarrow (13/2^+)$ | 4.9795 | 5 | $2.08 \times 10^{14}$ | $7.36 \times 10^{13}$ | 0.21 | 1.41 |
| $3/2^- \rightarrow (5/2)^-$ | 4.9171 | 2 | $1.40 \times 10^{14}$ | $2.58 \times 10^{15}$ | 0.32 | 0.04 |
| $3/2^- \rightarrow 3/2^-$ | 4.8603 | 0 | $2.45 \times 10^{14}$ | $1.23 \times 10^{13}$ | 0.18 | 8.46 |
| $3/2^- \rightarrow 1/2^-$ | 4.8358 | 2 | $5.31 \times 10^{14}$ | $1.03 \times 10^{14}$ | $8.40 \times 10^{-2}$ | 1.01 |
| $3/2^- \rightarrow 7/2^-$ | 4.8031 | 2 | $9.17 \times 10^{14}$ | $6.87 \times 10^{13}$ | $4.86 \times 10^{-2}$ | 1.51 |
| $3/2^- \rightarrow 5/2^+$ | 4.7647 | 1 | $1.38 \times 10^{15}$ | $1.03 \times 10^{15}$ | $3.23 \times 10^{-2}$ | $1.01 \times 10^{-1}$ |
| $3/2^- \rightarrow (5/2^-)$ | 4.7523 | 2 | $2.17 \times 10^{15}$ | $1.03 \times 10^{15}$ | $2.06 \times 10^{-2}$ | $1.01 \times 10^{-1}$ |
| $3/2^- \rightarrow (9/2^+)$ | 4.7211 | 3 | $5.29 \times 10^{15}$ | $6.87 \times 10^{16}$ | $8.43 \times 10^{-3}$ | $1.51 \times 10^{-3}$ |

| Transitions | Q | $\ell_{min}$ | $T_{cal.}$ | $T_{exp.}$ | $B_{cal.}$ | $B_{exp}$ |
|---|---|---|---|---|---|---|
| $3/2^- \to (3/2^-, 5/2^-)$ | 4.6889 | 0 | $4.49 \times 10^{15}$ | $1.29 \times 10^{16}$ | $9.93 \times 10^{-3}$ | $8.07 \times 10^{-3}$ |
| $3/2^- \to (3/2^+, 5/2^+)$ | 4.6293 | 1 | $1.45 \times 10^{16}$ | $3.44 \times 10^{16}$ | $3.08 \times 10^{-3}$ | $3.03 \times 10^{-3}$ |
| $3/2^- \to (7/2^+)$ | 4.5339 | 3 | $1.47 \times 10^{17}$ | $1.03 \times 10^{17}$ | $3.03 \times 10^{-4}$ | $1.01 \times 10^{-3}$ |
| $^{235}Np \to {}^{231}Pa$ | | | | | | |
| $5/2^+ \to 3/2^-$ | 5.2358 | 1 | $9.74 \times 10^{12}$ | $8.78 \times 10^{13}$ | 35.79 | 1.50 |
| $5/2^+ \to 1/2^-$ | 5.2266 | 3 | $2.03 \times 10^{13}$ | $6.58 \times 10^{14}$ | 17.17 | 0.20 |
| $5/2^+ \to 7/2^-$ | 5.1772 | 1 | $2.38 \times 10^{13}$ | $7.28 \times 10^{13}$ | 14.65 | 1.81 |
| $5/2^+ \to 5/2^+$ | 5.1516 | 0 | $3.13 \times 10^{13}$ | $2.48 \times 10^{12}$ | 11.14 | 53.26 |
| $5/2^+ \to 7/2^+$ | 5.1344 | 2 | $5.86 \times 10^{13}$ | $5.52 \times 10^{12}$ | 5.95 | 23.93 |
| $5/2^+ \to 3/2^+$ | 5.1335 | 2 | $5.94 \times 10^{13}$ | $4.89 \times 10^{14}$ | 5.87 | 0.27 |
| $5/2^+ \to (9/2^+)$ | 5.1242 | 2 | $6.86 \times 10^{13}$ | $2.19 \times 10^{13}$ | 5.08 | 6.03 |
| $5/2^+ \to 11/2^-$ | 5.0664 | 3 | $2.42 \times 10^{14}$ | $2.19 \times 10^{14}$ | 1.44 | 0.60 |
| $5/2^+ \to 5/2^+$ | 5.0523 | 0 | $1.49 \times 10^{14}$ | $1.14 \times 10^{13}$ | 2.34 | 11.59 |
| $5/2^+ \to 7/2^+$ | 4.9885 | 2 | $9.94 \times 10^{14}$ | $1.90 \times 10^{14}$ | 0.35 | 0.70 |
| $5/2^+ \to (9/2^+)$ | 4.9318 | 2 | $1.50 \times 10^{15}$ | $1.32 \times 10^{15}$ | 0.23 | 0.10 |
| $^{237}Np \to {}^{233}Pa$ | | | | | | |
| $5/2^+ \to 3/2^-$ | 5.0000 | 1 | $3.61 \times 10^{14}$ | $2.83 \times 10^{15}$ | 32.96 | 2.38 |
| $5/2^+ \to 7/2^-$ | 4.9429 | 1 | $9.16 \times 10^{14}$ | $2.78 \times 10^{15}$ | 12.99 | 2.42 |
| $5/2^+ \to 5/2^-$ | 4.9295 | 1 | $1.14 \times 10^{15}$ | $3.36 \times 10^{15}$ | 10.44 | 2.01 |
| $5/2^+ \to 5/2^+$ | 4.9135 | 0 | $1.32 \times 10^{15}$ | $1.42 \times 10^{14}$ | 9.01 | 47.46 |
| $5/2^+ \to 7/2^+$ | 4.3896 | 2 | $2.51 \times 10^{15}$ | $2.92 \times 10^{14}$ | 4.74 | 23.08 |
| $5/2^+ \to 9/2^+$ | 4.8910 | 2 | $2.74 \times 10^{15}$ | $7.28 \times 10^{14}$ | 4.34 | 9.26 |
| $5/2^+ \to (11/2^+)$ | 4.8668 | 4 | $9.21 \times 10^{15}$ | $3.56 \times 10^{17}$ | 1.29 | 0.02 |
| $5/2^+ \to (11/2^-)$ | 4.8367 | 3 | $9.71 \times 10^{15}$ | $1.13 \times 10^{16}$ | 1.23 | 0.60 |
| $5/2^+ \to (9/2^-)$ | 4.8209 | 3 | $1.27 \times 10^{16}$ | $1.26 \times 10^{16}$ | 0.94 | 0.53 |
| $5/2^+ \to 5/2^+$ | 4.7877 | 0 | $1.11 \times 10^{16}$ | $1.95 \times 10^{15}$ | 1.07 | 3.46 |

| Transitions | Q | $\ell_{min}$ | $T_{cal.}$ | $T_{exp.}$ | $B_{cal.}$ | $B_{exp}$ |
|---|---|---|---|---|---|---|
| $5/2^+ \to ?$ | 4.7820 | 0 | $1.23 \times 10^{16}$ | $2.26 \times 10^{16}$ | 0.97 | 0.30 |
| $5/2^+ \to 5/2^+$ | 4.7621 | 0 | $1.73 \times 10^{16}$ | $1.05 \times 10^{15}$ | 0.69 | 6.42 |
| $5/2^+ \to 5/2^-$ | 4.7429 | 1 | $2.73 \times 10^{16}$ | $2.11 \times 10^{17}$ | 0.44 | 0.03 |
| $5/2^+ \to (7/2^+)$ | 4.7203 | 2 | $5.13 \times 10^{16}$ | $1.82 \times 10^{16}$ | 0.23 | 0.37 |
| $5/2^+ \to 7/2^+$ | 4.6995 | 2 | $7.40 \times 10^{16}$ | $1.83 \times 10^{16}$ | 0.16 | 0.37 |
| $5/2^+ \to 9/2^+$ | 4.6341 | 2 | $2.39 \times 10^{17}$ | $1.93 \times 10^{17}$ | 0.05 | 0.03 |
| $5/2^+ \to 1/2^-$ | 4.9933 | 3 | $7.24 \times 10^{14}$ | $1.28 \times 10^{16}$ | 16.43 | 0.53 |
| $5/2^+ \to 1/2^+$ | 4.8308 | 2 | $7.57 \times 10^{15}$ | $1.13 \times 10^{16}$ | 1.57 | 0.60 |
| $5/2^+ \to ?$ | 4.7170 | 0 | $3.82 \times 10^{16}$ | $7.96 \times 10^{16}$ | 0.31 | 0.08 |
| $5/2^+ \to (9/2^+)$ | 4.6964 | 2 | $7.83 \times 10^{16}$ | $1.41 \times 10^{17}$ | 0.15 | 0.05 |
| $^{243}Am \to {}^{239}Np$ | | | | | | |
| $5/2^- \to 5/2^+$ | 5.6809 | 1 | $1.76 \times 10^{11}$ | $3.69 \times 10^{12}$ | 36.07 | 0.37 |
| $5/2^- \to 7/2^+$ | 5.6477 | 1 | $2.77 \times 10^{11}$ | $6.17 \times 10^{12}$ | 22.92 | 0.22 |
| $5/2^- \to 5/2^-$ | 5.6214 | 0 | $3.54 \times 10^{11}$ | $1.61 \times 10^{10}$ | 17.94 | 84.60 |
| $5/2^- \to 9/2^+$ | 5.6050 | 3 | $9.01 \times 10^{11}$ | $6.83 \times 10^{13}$ | 7.05 | 0.02 |
| $5/2^- \to 7/2^-$ | 5.5779 | 2 | $9.28 \times 10^{11}$ | $1.04 \times 10^{11}$ | 6.84 | 13.10 |
| $5/2^- \to 11/2^+$ | 5.5509 | 3 | $1.92 \times 10^{12}$ | $1.37 \times 10^{14}$ | 3.31 | 0.01 |
| $5/2^- \to 9/2^-$ | 5.5224 | 2 | $2.03 \times 10^{12}$ | $8.22 \times 10^{11}$ | 3.13 | 1.66 |
| $5/2^- \to 11/2^-$ | 5.4549 | 4 | $1.20 \times 10^{13}$ | $9.10 \times 10^{13}$ | $5.29 \times 10^{-1}$ | $1.50 \times 10^{-2}$ |
| $5/2^- \to 3/2^-$ | 5.4133 | 2 | $9.84 \times 10^{12}$ | $2.73 \times 10^{15}$ | $6.45 \times 10^{-1}$ | $4.99 \times 10^{-4}$ |
| $5/2^- \to 13/2^-$ | 5.3758 | 4 | $3.81 \times 10^{13}$ | $5.69 \times 10^{14}$ | $1.67 \times 10^{-1}$ | $2.39 \times 10^{-3}$ |
| $5/2^- \to (7/2^-)$ | 5.3565 | 2 | $2.28 \times 10^{13}$ | $1.05 \times 10^{15}$ | $2.78 \times 10^{-1}$ | $1.30 \times 10^{-3}$ |
| $5/2^- \to 1/2^+$ | 5.3485 | 3 | $3.63 \times 10^{13}$ | $1.37 \times 10^{17}$ | $1.75 \times 10^{-1}$ | $9.94 \times 10^{-6}$ |
| $5/2^- \to (5/2^-)$ | 5.3210 | 0 | $2.73 \times 10^{13}$ | $2.28 \times 10^{15}$ | $2.33 \times 10^{-1}$ | $5.97 \times 10^{-4}$ |
| $5/2^- \to 5/2^+$ | 5.3123 | 1 | $3.50 \times 10^{13}$ | $1.52 \times 10^{15}$ | $1.81 \times 10^{-1}$ | $8.96 \times 10^{-4}$ |
| $5/2^- \to 3/2^+$ | 5.3100 | 1 | $3.62 \times 10^{13}$ | $4.55 \times 10^{15}$ | $1.75 \times 10^{-1}$ | $2.99 \times 10^{-4}$ |
| $5/2^- \to 15/2^-$ | 5.2853 | 6 | $4.83 \times 10^{13}$ | $1.95 \times 10^{15}$ | $1.31 \times 10^{-1}$ | $6.99 \times 10^{-4}$ |

| Transitions | Q | $\ell_{min}$ | $T_{cal.}$ | $T_{exp.}$ | $B_{cal.}$ | $B_{exp}$ |
|---|---|---|---|---|---|---|
| 5/2⁻→? | 5.2627 | 0 | 6.61x10¹³ | 4.27x10¹⁵ | 9.61x10⁻² | 3.19x10⁻⁴ |
| 5/2⁻→(11/2⁻) | 5.2468 | 4 | 2.66x10¹⁴ | 3.41x10¹⁵ | 2.39x10⁻² | 3.99x10⁻⁴ |
| 5/2⁻→9/2⁺ | 5.2284 | 3 | 2.25x10¹⁴ | 3.41x10¹⁵ | 2.82x10⁻² | 3.99x10⁻⁴ |
| 5/2⁻→7/2⁺ | 5.2212 | 1 | 1.41x10¹⁴ | 3.41x10¹⁵ | 4.50x10⁻² | 3.99x10⁻⁴ |
| 5/2⁻→(9/2⁻) | 5.1947 | 2 | 2.69x10¹⁴ | 9.75x10¹⁵ | 2.36x10⁻² | 1.40x10⁻⁴ |
| 5/2⁻→(5/2⁻) | 5.1348 | 0 | 4.86x10¹⁴ | 1.37x10¹⁶ | 1.31x10⁻² | 9.94x10⁻⁵ |
| 5/2⁻→5/2⁻ | 4.9589 | 0 | 8.54x10¹⁵ | 1.95x10¹⁵ | 7.43x10⁻⁴ | 6.99x10⁻⁴ |
| 5/2⁻→7/2⁻ | 4.9252 | 2 | 2.13x10¹⁶ | 1.59x10¹⁶ | 2.98x10⁻⁴ | 8.57x10⁻⁵ |
| 5/2⁻→9/2⁻ | 4.8811 | 2 | 4.51x10¹⁶ | 2.73x10¹⁷ | 1.41x10⁻⁴ | 4.99x10⁻⁶ |
| $^{241}$Am → $^{237}$Np | | | | | | |
| 5/2⁻→5/2⁺ | 5.4819 | 1 | 2.67x10¹² | 1.45x10¹⁴ | 43.94 | 0.16 |
| 5/2⁻→7/2⁺ | 5.4508 | 1 | 4.18x10¹² | 1.45x10¹⁴ | 28.07 | 0.16 |
| 5/2⁻→5/2⁻ | 5.4072 | 0 | 7.03x10¹² | 2.67x10¹¹ | 16.69 | 87.12 |
| 5/2⁻→7/2⁻ | 5.3641 | 2 | 1.89x10¹³ | 2.08x10¹² | 6.21 | 11.18 |
| 5/2⁻→9/2⁻ | 5.3089 | 2 | 4.33x10¹³ | 1.71x10¹³ | 2.71 | 1.36 |
| 5/2⁻→(11/2⁻) | 5.2411 | 4 | 2.69x10¹⁴ | 2.91x10¹⁵ | 4.36x10⁻¹ | 7.99x10⁻³ |
| 5/2⁻→(5/2⁺) | 5.2149 | 1 | 1.44x10¹⁴ | 4.65x10¹⁵ | 8.15x10⁻¹ | 5.00x10⁻³ |
| 5/2⁻→(13/2⁻) | 5.1645 | 4 | 8.83x10¹⁴ | 1.55x10¹⁶ | 1.33x10⁻¹ | 1.50x10⁻³ |
| 5/2⁻→(5/2⁻) | 5.1569 | 0 | 3.18x10¹⁴ | 1.55x10¹⁶ | 3.69x10⁻¹ | 1.50x10⁻³ |
| 5/2⁻→? | 5.1349 | 0 | 4.51x10¹⁴ | 2.91x10¹⁶ | 2.60x10⁻¹ | 7.99x10⁻⁴ |
| 5/2⁻→? | 5.1229 | 0 | 5.46x10¹⁴ | 2.91x10¹⁶ | 2.15x10⁻¹ | 7.99x10⁻⁴ |
| 5/2⁻→? | 5.0709 | 0 | 1.26x10¹⁵ | 6.84x10¹⁶ | 9.31x10⁻² | 3.40x10⁻⁴ |
| 5/2⁻→? | 5.0549 | 0 | 1.63x10¹⁵ | 1.29x10¹⁷ | 7.20x10⁻² | 1.80x10⁻⁴ |
| $^{247}$Bk → $^{243}$Am | | | | | | |
| (3/2⁻)→5/2⁻ | 5.9345 | 2 | 7.33x10¹⁰ | 7.92x10¹¹ | 42.37 | 5.54 |
| (3/2⁻)→7/2⁻ | 5.8935 | 2 | 1.26x10¹¹ | 1.01x10¹² | 24.65 | 4.35 |

| Transitions | Q | $\ell_{min}$ | $T_{cal.}$ | $T_{exp.}$ | $B_{cal.}$ | $B_{exp}$ |
|---|---|---|---|---|---|---|
| $(3/2^-)\to 5/2^+$ | 5.3850 | 1 | $1.76\times10^{11}$ | $2.56\times10^{11}$ | 17.65 | 17.15 |
| $(3/2^-)\to 7/2^+$ | 5.8265 | 3 | $4.32\times10^{11}$ | $3.35\times10^{11}$ | 7.19 | 13.10 |
| $(3/2^-)\to (9/2^+)$ | 5.7925 | 3 | $6.82\times10^{11}$ | $7.92\times10^{11}$ | 4.55 | 5.54 |
| $(3/2^-)\to (11/2^+)$ | 5.7475 | 5 | $3.82\times10^{12}$ | $1.09\times10^{13}$ | 0.81 | 0.40 |
| $(3/2^-)\to (3/2^-)$ | 5.6685 | 0 | $1.89\times10^{12}$ | $9.68\times10^{10}$ | 1.64 | 45.35 |
| $(3/2^-)\to (5/2^-)$ | 5.6365 | 2 | $4.18\times10^{12}$ | $6.22\times10^{11}$ | 0.74 | 7.06 |
| $(3/2^-)\to (7/2^-)$ | 5.5905 | 2 | $8.04\times10^{12}$ | $2.90\times10^{12}$ | 0.39 | 1.51 |
| $^{249}$Bk $\to$ $^{245}$Am | | | | | | |
| $7/2^+\to (5/2)^+$ | 5.5692 | 2 | $1.01\times10^{13}$ | $3.95\times10^{13}$ | 29.73 | 4.88 |
| $7/2^+\to (7/2^+)$ | 5.5492 | 0 | $9.58\times10^{12}$ | $2.56\times10^{12}$ | 31.34 | 75.29 |
| $7/2^+\to (9/2^+)$ | 5.5210 | 2 | $2.03\times10^{13}$ | $1.19\times10^{13}$ | 14.79 | 16.20 |
| $7/2^+\to (11/2^+)$ | 5.4822 | 2 | $3.59\times10^{13}$ | $1.27\times10^{14}$ | 8.36 | 1.52 |
| $7/2^+\to ?$ | 5.4522 | 0 | $3.96\times10^{13}$ | $3.95\times10^{15}$ | 7.58 | 0.05 |
| $7/2^+\to (9/2^-)$ | 5.4442 | 1 | $5.00\times10^{13}$ | $2.58\times10^{15}$ | 6.00 | 0.07 |
| $7/2^+\to (11/2^-)$ | 5.3784 | 3 | $2.36\times10^{14}$ | $1.91\times10^{15}$ | 1.27 | 0.10 |
| $7/2^+\to ?$ | 5.2792 | 0 | $5.48\times10^{14}$ | $5.03\times10^{15}$ | 0.55 | 0.04 |
| $7/2^+\to 7/2^+$ | 5.2420 | 0 | $9.81\times10^{14}$ | $1.06\times10^{14}$ | 0.31 | 1.82 |
| $7/2^+\to 9/2^+$ | 5.1712 | 2 | $4.25\times10^{15}$ | $4.77\times10^{15}$ | 0.07 | 0.04 |
| $^{247}$Es $\to$ $^{243}$Bk | | | | | | |
| $(7/2^+)\to (7/2^+)$ | 7.4940 | 0 | $1.59\times10^{4}$ | $4.55\times10^{3}$ | 59.74 | 85.96 |
| $(7/2^+)\to (9/2^+)$ | 7.445 | 2 | $3.65\times10^{4}$ | $3.25\times10^{4}$ | 26.02 | 12.03 |
| $(7/2^+)\to (11/2^+)$ | 7.3820 | 2 | $6.67\times10^{4}$ | $1.95\times10^{5}$ | 14.24 | 2.01 |
| $^{251}$Es $\to$ $^{247}$Bk | | | | | | |
| $(3/2^-)\to (3/2^-)$ | 6.6420 | 0 | $9.01\times10^{7}$ | $2.90\times10^{7}$ | 43.05 | 81.02 |
| $(3/2^-)\to (5/2^-)$ | 6.6120 | 2 | $1.80\times10^{8}$ | $2.53\times10^{8}$ | 21.55 | 9.29 |

| Transitions | Q | $\ell_{min}$ | $T_{cal.}$ | $T_{exp.}$ | $B_{cal.}$ | $B_{exp}$ |
|---|---|---|---|---|---|---|
| (3/2⁻)→(7/2⁺) | 6.6010 | 3 | 2.87x10⁸ | 6.99x10⁸ | 13.52 | 3.36 |
| (3/2⁻)→(7/2⁻) | 6.5710 | 2 | 2.87x10⁸ | 7.92x10⁸ | 13.52 | 2.97 |
| (3/2⁻)→(9/2⁺) | 6.5590 | 3 | 4.64x10⁸ | 6.99x10⁸ | 8.36 | 3.36 |
| ²⁵³Es → ²⁴⁹Bk | | | | | | |
| 7/2⁺→7/2⁺ | 6.7851 | 0 | 1.71x10⁷ | 1.97x10⁶ | 28.26 | 89.80 |
| 7/2⁺→(3/2⁻) | 6.7763 | 3 | 3.78x10⁷ | 2.21x10⁸ | 12.78 | 0.80 |
| 7/2⁺→(5/2⁻) | 6.7455 | 1 | 2.97x10⁷ | 2.53x10⁸ | 16.27 | 0.70 |
| 7/2⁺→9/2⁺ | 6.7433 | 2 | 3.84x10⁷ | 2.68x10⁷ | 12.58 | 6.60 |
| 7/2⁺→(7/2⁻) | 6.7025 | 1 | 4.78x10⁷ | 2.49x10⁸ | 10.11 | 0.71 |
| 7/2⁺→11/2⁺ | 6.6913 | 2 | 6.84x10⁷ | 2.08x10⁸ | 7.06 | 0.85 |
| 7/2⁺→(9/2⁻) | 6.6474 | 1 | 8.86x10⁷ | 6.80x10⁸ | 5.45 | 0.26 |
| 7/2⁺→13/2⁺ | 6.6293 | 2 | 1.37x10⁸ | 2.08x10⁹ | 3.53 | 8.51x10⁻² |
| 7/2⁺→(11/2⁻) | 6.5805 | 3 | 3.36x10⁸ | 2.90x10⁹ | 1.44 | 6.10x10⁻² |
| 7/2⁺→(15/2⁺) | 6.5558 | 4 | 6.98x10⁸ | 1.36x10¹⁰ | 6.92x10⁻¹ | 1.30x10⁻² |
| 7/2⁺→(13/2⁻) | 6.5019 | 3 | 8.32x10⁸ | 2.16x10¹⁰ | 5.81x10⁻¹ | 8.19x10⁻³ |
| 7/2⁺→(17/2⁺) | 6.4721 | 6 | 5.92x10⁹ | 4.42x10¹¹ | 8.16x10⁻² | 4.00x10⁻⁴ |
| 7/2⁺→(15/2⁻) | 6.4122 | 5 | 6.39x10⁹ | 2.21x10¹¹ | 7.56x10⁻² | 8.01x10⁻⁴ |
| 7/2⁺→(5/2)⁺ | 6.3959 | 2 | 2.06x10⁹ | 3.93x10⁹ | 2.35x10⁻¹ | 4.50x10⁻² |
| 7/2⁺→(3/2⁺) | 6.3745 | 2 | 2.66x10⁹ | 1.47x10¹² | 1.82x10⁻¹ | 1.20x10⁻⁴ |
| 7/2⁺→(5/2⁺) | 6.3639 | 2 | 3.02x10⁹ | 1.18x10¹¹ | 1.60x10⁻¹ | 1.50x10⁻³ |
| 7/2⁺→(7/2)⁺ | 6.3562 | 0 | 2.35x10⁹ | 4.53x10⁹ | 2.06x10⁻¹ | 3.91x10⁻² |
| 7/2⁺→(9/2⁺) | 6.3102 | 2 | 5.78x10⁹ | 1.18x10¹⁰ | 8.36x10⁻² | 1.50x10⁻² |
| 7/2⁺→(7/2⁺) | 6.2662 | 0 | 7.02x10⁹ | 2.27x10¹¹ | 6.88x10⁻² | 7.79x10⁻⁴ |
| 7/2⁺→(11/2⁺) | 6.2431 | 2 | 1.32x10¹⁰ | 5.20x10¹⁰ | 3.66x10⁻² | 3.40x10⁻³ |
| 7/2⁺→(3/2⁻) | 6.2270 | 3 | 2.25x10¹⁰ | 7.07x10¹¹ | 2.15x10⁻² | 2.50x10⁻⁴ |
| 7/2⁺→(1/2⁻) | 6.2159 | 3 | 2.59x10¹⁰ | 5.90x10¹² | 1.87x10⁻² | 3.00x10⁻⁵ |
| 7/2⁺→(13/2⁺) | 6.1875 | 4 | 5.73x10¹⁰ | 4.42x10¹¹ | 8.43x10⁻³ | 4.00x10⁻⁴ |

| Transitions | Q | $\ell_{min}$ | $T_{cal.}$ | $T_{exp.}$ | $B_{cal.}$ | $B_{exp}$ |
|---|---|---|---|---|---|---|
| $7/2^+ \to (7/2^-)$ | 6.1785 | 1 | $2.34 \times 10^{10}$ | $6.10 \times 10^{11}$ | $2.07 \times 10^{-2}$ | $2.90 \times 10^{-4}$ |
| $7/2^+ \to (5/2^-)$ | 6.1601 | 1 | $2.95 \times 10^{10}$ | $9.83 \times 10^{11}$ | $1.64 \times 10^{-2}$ | $1.80 \times 10^{-4}$ |
| $7/2^+ \to ?$ | 6.1161 | 0 | $4.60 \times 10^{10}$ | $2.95 \times 10^{12}$ | $1.05 \times 10^{-2}$ | $6.00 \times 10^{-5}$ |
| $7/2^+ \to (11/2^-)$ | 6.0851 | 3 | $1.35 \times 10^{11}$ | $1.18 \times 10^{12}$ | $3.58 \times 10^{-3}$ | $1.50 \times 10^{-4}$ |
| $7/2^+ \to ?$ | 6.0761 | 0 | $7.68 \times 10^{10}$ | $4.42 \times 10^{12}$ | $6.29 \times 10^{-3}$ | $4.00 \times 10^{-5}$ |
| $7/2^+ \to ?$ | 6.0511 | 0 | $1.06 \times 10^{11}$ | $6.55 \times 10^{12}$ | $4.56 \times 10^{-3}$ | $2.70 \times 10^{-5}$ |
| $7/2^+ \to ?$ | 5.8528 | 0 | $1.48 \times 10^{12}$ | $2.21 \times 10^{12}$ | $3.26 \times 10^{-4}$ | $8.01 \times 10^{-5}$ |
| $^{255}Es \to {}^{251}Bk$ | | | | | | |
| $(7/2^+) \to (7/2^+)$ | 6.8071 | 0 | $1.25 \times 10^{7}$ | $4.91 \times 10^{7}$ | 57.37 | 87.73 |
| $(7/2^+) \to (9/2^+)$ | 6.7720 | 2 | $2.61 \times 10^{7}$ | $4.41 \times 10^{8}$ | 27.47 | 9.77 |
| $(7/2^+) \to (11/2^+)$ | 6.7180 | 2 | $4.73 \times 10^{7}$ | $1.72 \times 10^{9}$ | 15.16 | 2.50 |
| $^{255}Md \to {}^{251}Es$ | | | | | | |
| $(7/2^-) \to (7/2^-)$ | 7.4919 | 0 | $1.02 \times 10^{5}$ | $2.19 \times 10^{4}$ | 72.21 | 94.87 |
| $(7/2^-) \to 9/2^-$ | 7.4295 | 2 | $2.65 \times 10^{5}$ | $4.05 \times 10^{5}$ | 27.79 | 5.13 |
| $^{257}Md \to {}^{253}Es$ | | | | | | |
| $(7/2^-) \to (7/2^+)$ | 7.6049 | 1 | $3.64 \times 10^{4}$ | $3.68 \times 10^{7}$ | 37.79 | 0.37 |
| $(7/2^-) \to 9/2^+$ | 7.5586 | 1 | $5.63 \times 10^{4}$ | $4.97 \times 10^{7}$ | 24.43 | 0.27 |
| $(7/2^-) \to 11/2^+$ | 7.5249 | 3 | $1.39 \times 10^{5}$ | $1.32 \times 10^{8}$ | 9.90 | 0.10 |
| $(7/2^-) \to 3/2^-$ | 7.4989 | 2 | $1.26 \times 10^{5}$ | $9.94 \times 10^{7}$ | 10.92 | 0.14 |
| $(7/2^-) \to 5/2^-$ | 7.4659 | 2 | $1.73 \times 10^{5}$ | $5.23 \times 10^{7}$ | 7.95 | 0.26 |
| $(7/2^-) \to 7/2^-$ | 7.4236 | 0 | $1.84 \times 10^{5}$ | $6.62 \times 10^{7}$ | 7.48 | 0.20 |
| $(7/2^-) \to 7/2^-$ | 7.2335 | 0 | $1.22 \times 10^{6}$ | $1.42 \times 10^{5}$ | 1.13 | 95.32 |
| $(7/2^-) \to 9/2^-$ | 7.1699 | 2 | $3.30 \times 10^{6}$ | $4.06 \times 10^{6}$ | 0.42 | 3.33 |

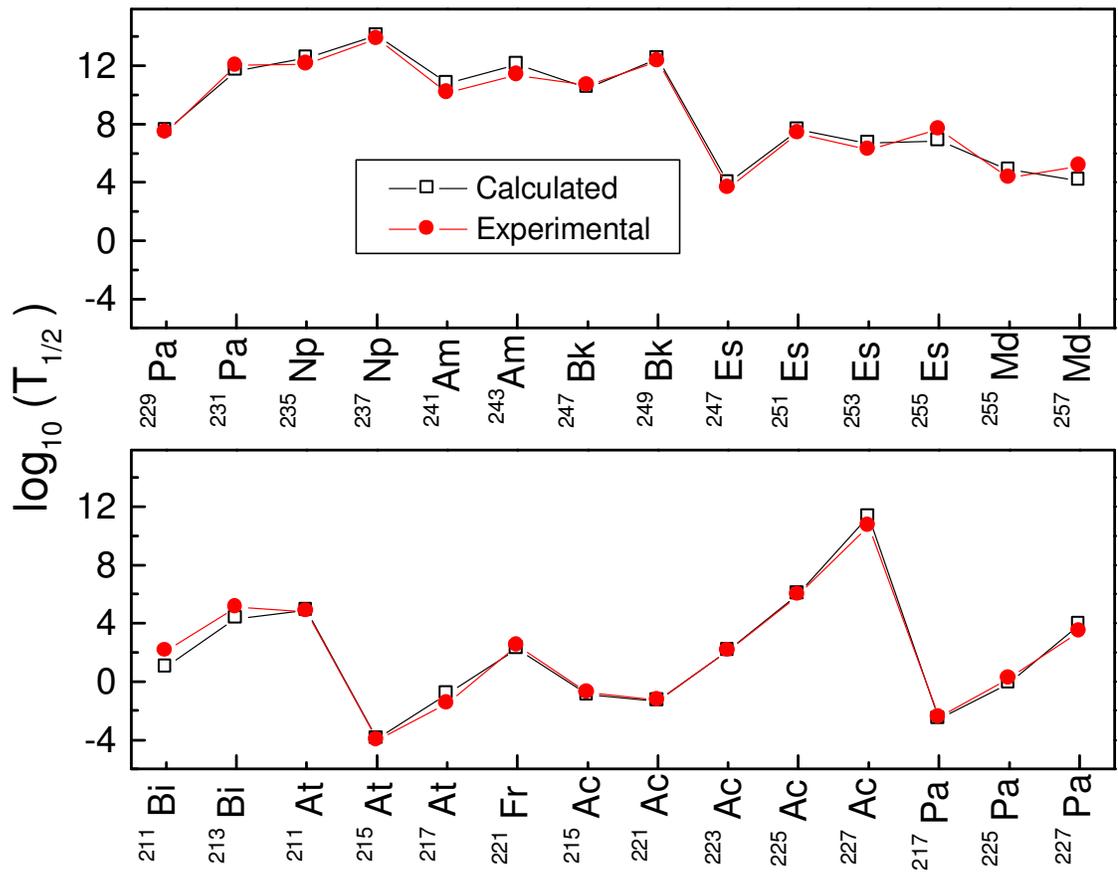

Fig. 1. The comparison of calculated total half life of various nuclei with the corresponding experimental values.

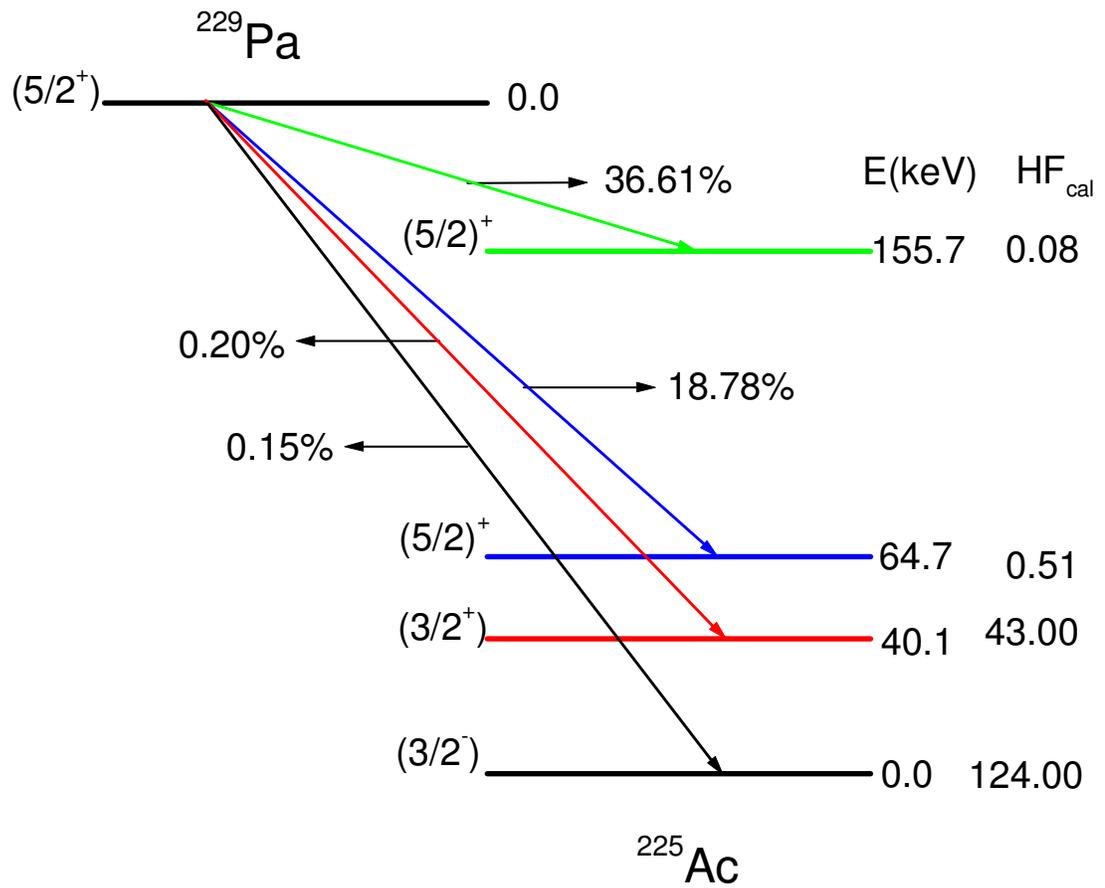

Fig. 2. A schematic representation of α-decay from ground state of $^{229}$Pa to few levels of $^{224}$Ac

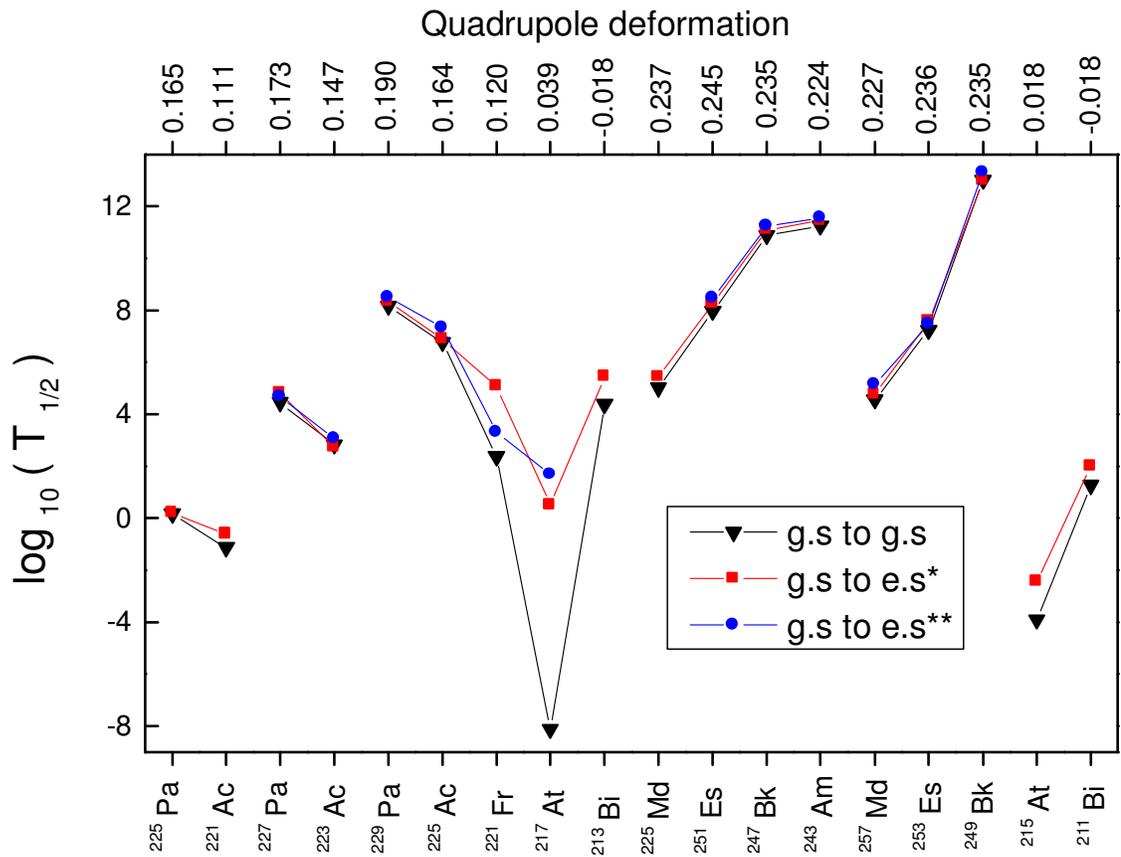

Fig 3. Plot connecting the logarithmic half lives and deformation values for various parent nuclei. e.s* and e.s** represents the first excited state and second excited state respectively.

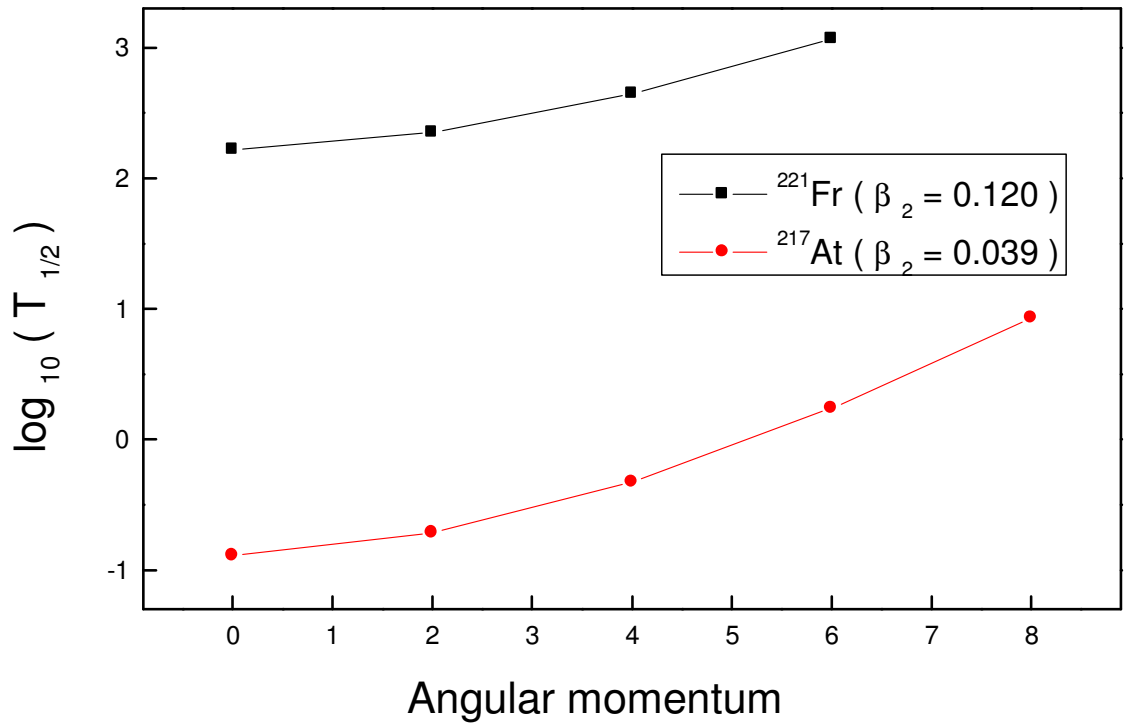

Fig 4. Plot connecting logarithmic half lives and angular momentum for $^{221}$Fr and $^{217}$At.

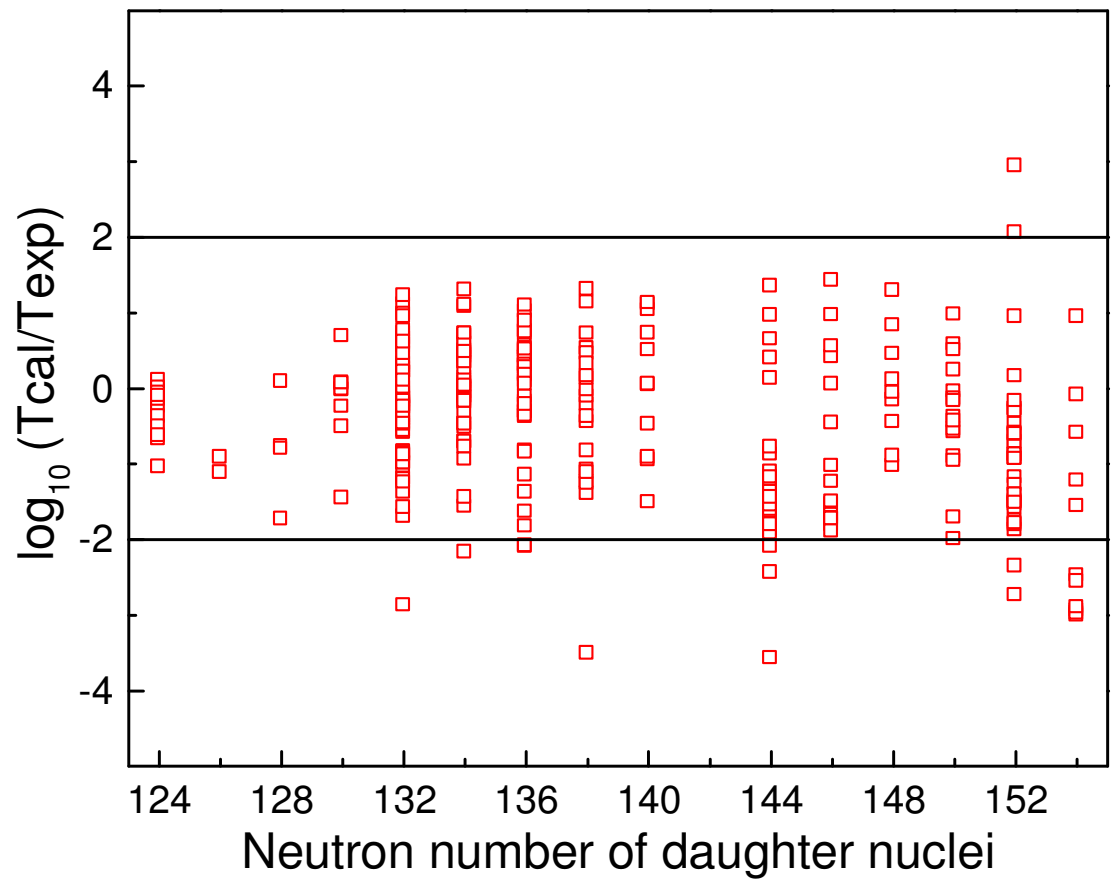

Fig. 5. The deviation of calculated $T_{1/2}$ values with the corresponding experimental data for all transitions.